\documentclass[aps,apl,preprint,groupedaddress]{revtex4-1}

\setcitestyle{super}

\usepackage{graphicx}

\usepackage{amsfonts, amsmath, amssymb,latexsym}

\usepackage{dcolumn}

\usepackage{bm}

\usepackage{gensymb}

\usepackage{xcolor}

\newcommand{\dif}{\mathrm{d}}

\begin{document}

\title{Comparison of Single-Ion Molecular Dynamics in Common Solvents}

\author{A. Muralidharan, L. R. Pratt} 
\affiliation{Department of Chemical and Biomolecular Engineering, Tulane University, New Orleans, LA 70118, USA}
\author{M. I. Chaudhari, S. B. Rempe} 
\affiliation{Sandia National Laboratories, Center for Biological and Engineering Sciences, Albuquerque, 87185, USA}

\date{\today}

\begin{abstract}
Laying a basis for molecularly specific theory for the mobilities of
ions in solutions of practical interest, we report a broad survey of
velocity autocorrelation functions (VACFs) of Li$^+$ and PF$_6{}^-$ ions
in water, ethylene carbonate, propylene carbonate, and acetonitrile
solutions. We extract the memory function, $\gamma(t)$, which
characterizes the random forces governing the mobilities of ions. We
provide comparisons, controlling for electrolyte concentration and
ion-pairing, for van~der~Waals attractive interactions and  solvent
molecular characteristics. For the heavier ion (PF$_6{}^-$), velocity
relaxations are all similar: negative tail relaxations for the VACF and
a clear second relaxation for $\gamma\left(t\right)$, observed
previously also for other molecular ions and with \emph{n}-pentanol as
solvent. For the light Li$^+$ ion, short time-scale oscillatory behavior
masks simple, longer time-scale relaxation of $\gamma\left(t\right)$.
But the corresponding analysis of the \emph{solventberg} 
Li$^+\left(\mathrm{H}_2\mathrm{O}\right)_4$ does conform to the standard
picture set by all the PF$_6{}^-$ results. 

\end{abstract}

\maketitle \section{Introduction} 
Here we report molecular dynamics results for
single-ion dynamics in liquid solutions, including aqueous solutions. We provide
comparisons controlling for the effects of solvent molecular characteristics,
electrolyte concentration, and van~der~Waals attractive forces. We choose
LiPF$_6$ for our study because of its importance, with ethylene carbonate (EC),
to lithium ion batteries. But our comparisons include several solvents of
experimental interest, specifically water, EC, propylene carbonate (PC), and
acetonitrile (ACN). We obtain the memory function $\gamma(t)$, defined
below,\cite{forster1975hydrodynamic} which characterizes the random forces
governing the mobilities of ions in these solvents. 

\begin{figure}
\includegraphics[width=3.25in]{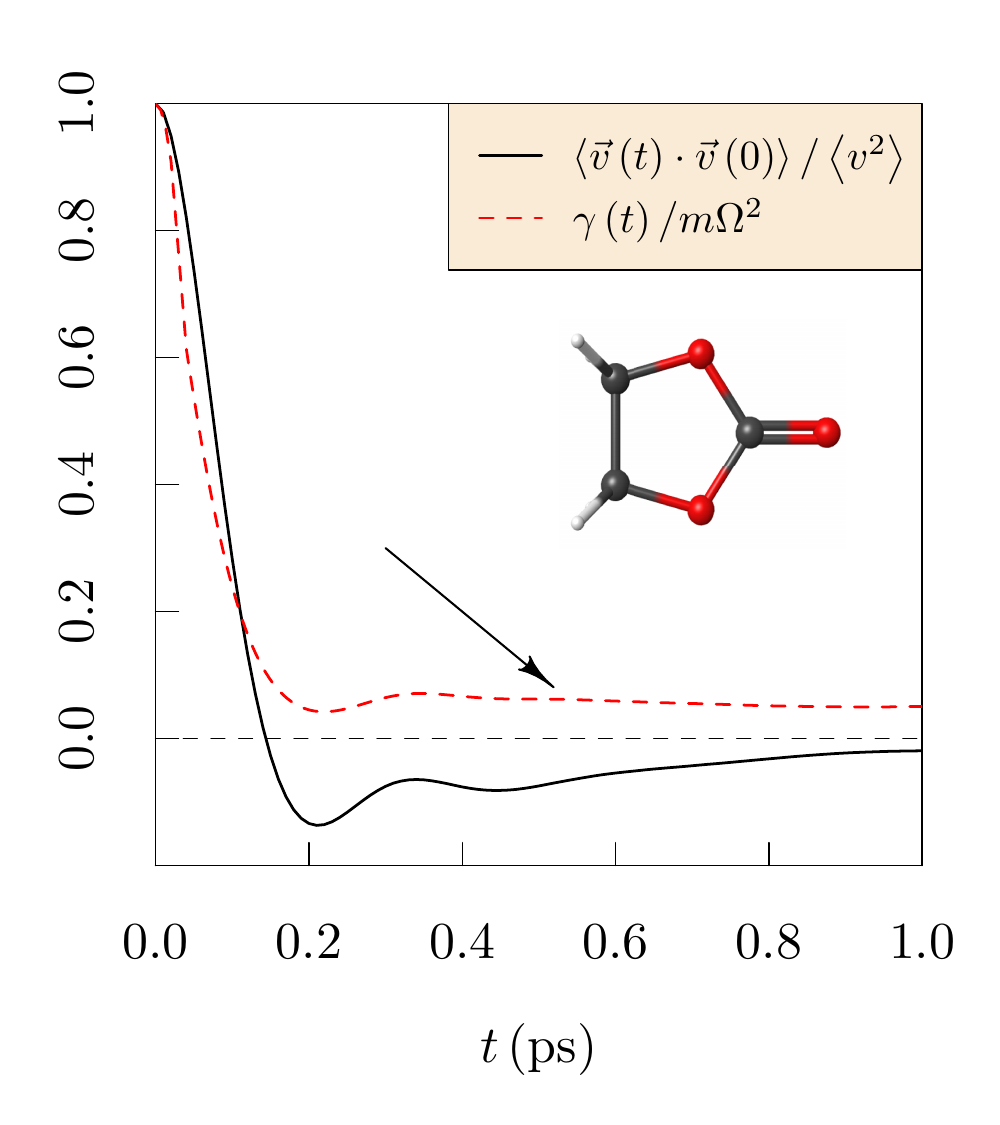}
\caption{ Velocity autocorrelation function and friction kernel
$\gamma\left(t \right)$ defined with Eq.~\eqref{eq:gle}, for the
center-of-mass of ethylene carbonate in neat liquid ethylene carbonate. 
 See also Ref.~\citenum{you2014role}.  The arrow indicates the
second-relaxation feature that is the primary phenomenon for these
studies.}
\label{fig:EC-PC} \end{figure}

A specific motivation for this work is the direct observation
\cite{zhu2012pairing} that $\gamma(t)$ relaxes  on time scales longer
than the direct collisional time-scale, behavior that was anticipated
years earlier in the context of \emph{dielectric
friction.}\cite{wolynes1978molecular} Nevertheless, this longer
time-scale relaxation is not limited to ionic interactions
(FIG.~\ref{fig:EC-PC}).\cite{you2014role} The results and comparisons
below provide a basis for molecularly specific theory for the 
mobilities in liquid mixtures of highly asymmetric species, as are
electrolyte solutions of practical
interest.\cite{Zwanzig:1970va,Metiu:1977wz,Gaskell:2001cq,Balucani:1999ec}

\section{\label{sec:level1}Methods}
We perform simulations (Table \ref{Table:MD_system}) of dilute and 1M
solutions of LiPF$_6$ using the GROMACS molecular dynamics package with
periodic boundary conditions. A Nose-Hoover
thermostat\cite{Nose,Hoover} and a Parrinello-Rahman\cite{barostat}
barostat were utilized to achieve equilibration in the $NpT$ ensemble at
300~K and 1~atm pressure. A 10~ns simulation was carried out for aging,
then a separate 1~ns simulation with a sampling rate of 1~fs was carried
out to calculate the velocity autocorrelation and the friction kernel.

\begin{table}[]
\centering
\begin{tabular}{|c|c|c|c|c|c|}
\hline
 & \textbf{Ions}       & \textbf{Water} & \textbf{ACN} & \textbf{EC}  & \textbf{PC}  \\ \hline
\textbf{1M}            & 32 Li$^+$ $+$ 32 PF$_6{}^-$ & 1776  & 613 & 480 & 378 \\ 
\textbf{dilute}        & 1 Li$^+$ / 1 PF$_6{}^-$   & 999   & 449 & 249 & 249 \\ 
\hline
\end{tabular}%
\caption{System sizes in dilute and concentrated
solutions of 1M LiPF$_6$ in several solvents}
\label{Table:MD_system}
\end{table} 
 
\begin{figure}
\includegraphics[width=3.5in]{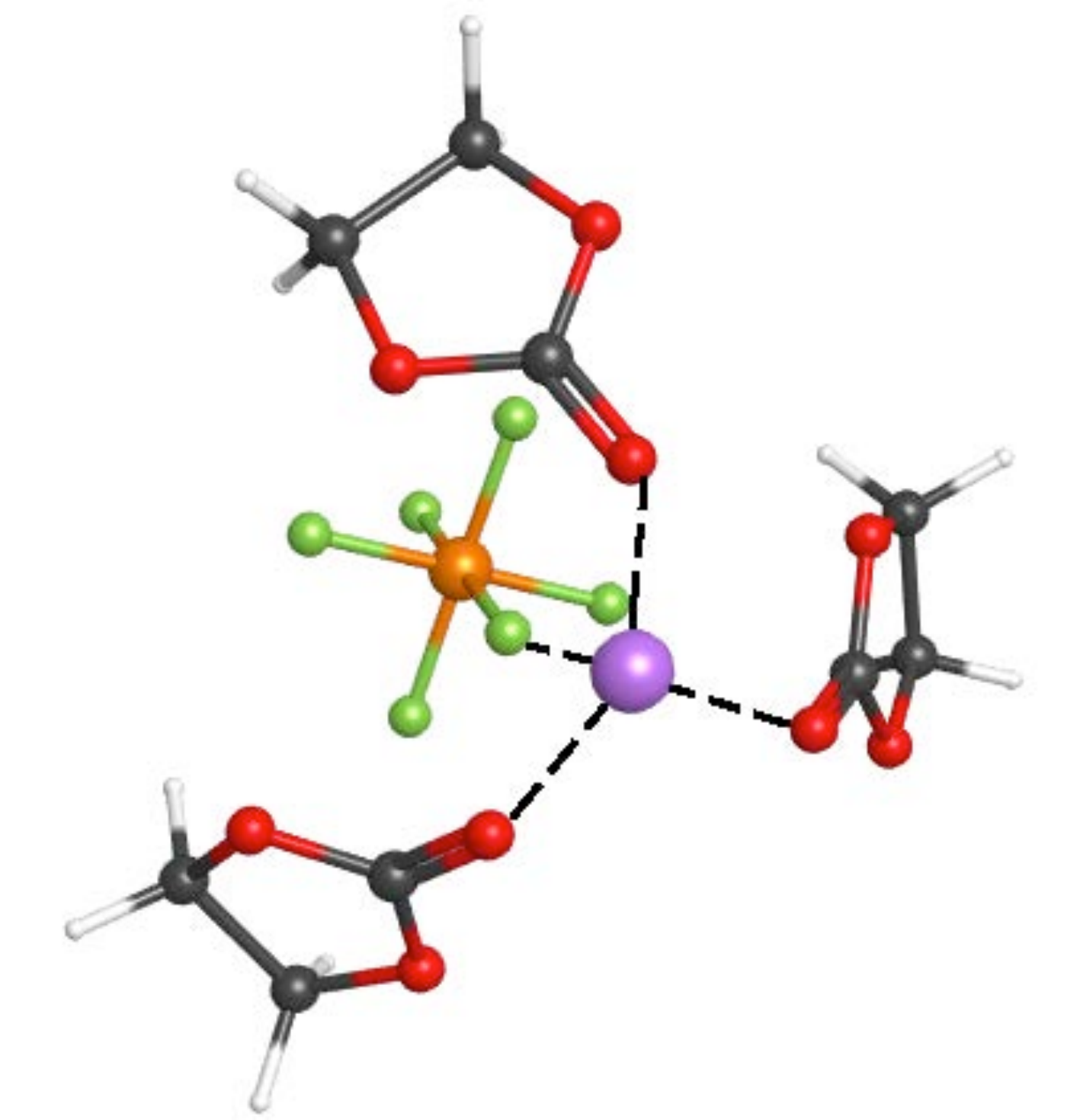}
\caption{Optimized (EC)$_3$Li$^+$\ldots PF$_6{}^-$ cluster,
characterizing ion-paired structures of LiPF$_6$ in ethylene carbonate
solvent. Gaussian09\cite{g09} was used for these electronic
structure calculations at the Hartee-Fock level with
{\tt{6-311+G(2d,p)}} basis set.  Initial structures were sampled from MD
simulations. }
\label{fig:cluster} \end{figure}

\subsection{Forcefield parameters and adjustments}
The interactions were modeled following the OPLS-AA
forcefield\cite{oplsaa} with parameters as indicated below for bonded
and non-bonded interactions. Li$^+$ parameters were obtained from
Soetens, \emph{et al.}\cite{soetens1998molecular} Partial charges of EC
and PC were scaled \cite{chaudhari2016scaling} to match transport
properties of Li$^+$ with experiment. In the case of acetonitrile and
water, standard OPLS-AA and SPC/E parameters were
used.\cite{berendsen1987missing} 

The PF$_6{}^-$ ions were described initially with parameters from
Sharma, \emph{et al.}\cite{sharma2016effect}  In initial MD trials,
however, we observed PF$_6{}^-$ ions that deviated significantly from
octahedral geometries, particularly in the case of 1M LiPF$_6$ in EC,
where substantial ion-pairing was observed. These PF$_6{}^-$ displayed
extreme bending of the axial F-P-F bond angles. 

The possibility of exotic non-octahedral PF$_6{}^-$ configurations in
ion-paired (EC)$_3$Li$^+$\ldots PF$_6{}^-$ clusters was investigated
with electronic structure calculations. Gaussian09
calculations\cite{g09} employed the Hartee-Fock approximation with the
{\tt{6-311+G(2d,p)}} basis set. Initial configurations were sampled
from MD observations.  The stable and lowest-energy clusters obtained
were consistent with octahedral PF$_6{}^-$ geometries
(FIG.~\ref{fig:cluster}). We therefore increased the axial F-P-F
(180$\degree$) bond-angle parameter by a factor of four in further MD
calculations. The modified forcefield parameters for PF$_6{}^-$ are
provided with supplementary information.

\begin{figure*}
\includegraphics[width=7in]{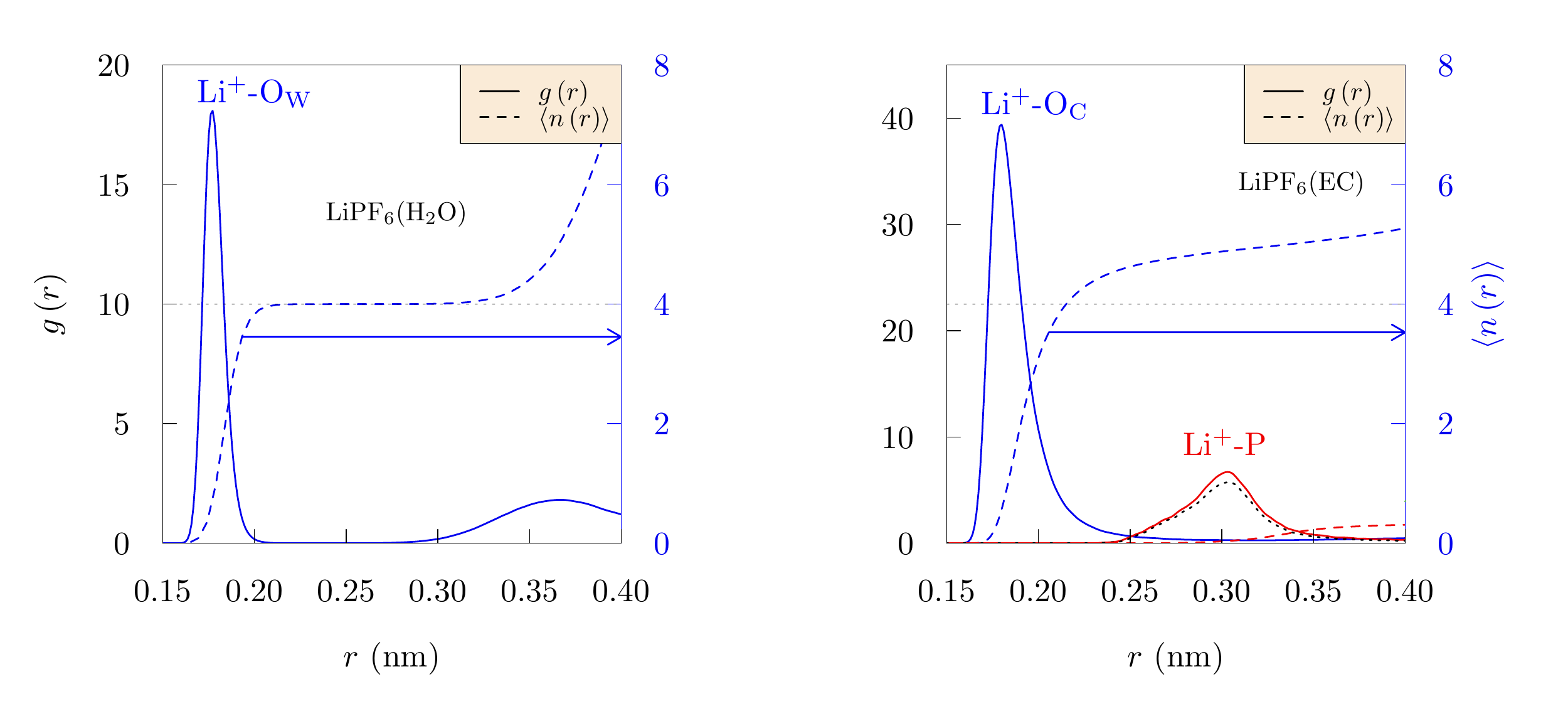}
\caption{Radial distribution functions (solid curves, left axes), and
the corresponding running coordination numbers (dashed, right axes). 
The
right-pointing arrows indicate the axes for the coordination numbers
$\langle{n(r)}\rangle$. Left panel: 1M LiPF$_6$ in water with
$N_\mathrm{W}$ = 1776 water molecules. The extended
$\langle{n(r)}\rangle = 4$ plateau shows a distinct inner shell with
that occupancy. Right panel: 1M LiPF$_6$ in ethylene carbonate with
$N_\mathrm{EC}$ = 480 EC molecules. In contrast to the water case, a
P atom is localized with the O$_\mathrm{C}$ inner shell.
The black-dotted curve is $g\left(r\right)\exp\left\{
-\langle{n(r)}\rangle \right\}$, the Fuoss/Poisson
approximation\cite{zhu2011generalizations} to the distribution of the
nearest P atom to a Li$^+$  ion, supporting 
Li$^+ \cdots$ PF$_6{}^-$ ion pairing at this concentration.}
\label{fig:rdfcn_compare2} \end{figure*}

\subsection{\label{sec:level2B}Solution structure}
For Li$^+$ in water, the oxygen coordination number is
4,\cite{rempe2000hydration,alam2011computing,mason2015neutron} with the
inner-shell O atoms positioned  at 0.18 nm. Similar Li$^+$ coordination
is observed in 1M solutions of LiPF$_6$ in PC and ACN. 

In the case of 1M solutions in EC, the nearest Li-P peak centered at 
0.33 nm (FIG.~\ref{fig:rdfcn_compare2}) indicates distinct but modest ion-pairing with
PF$_6{}^-$ at this concentration.   The
Fuoss/Poisson approximation\cite{zhu2011generalizations} is accurate
here and that further supports the ion-pairing picture.  Reflecting 
F atom penetration of the natural EC inner shell (FIG.~\ref{fig:cluster}), 
the Li$^+$-O atom inner shell distribution is broader 
in EC than in water.
 
We re-emphasize that previous work\cite{chaudhari2016scaling} scaled partial 
charges of the solvent EC molecules to match \emph{ab initio} and
experimental results for Li$^+$ solvation and dynamics.  
Nevertheless, van~der~Waals interactions are a primary 
concern for description of realistic ion-pairing. 

\subsection{The friction kernel}
We define the friction kernel $\gamma\left(t\right)$ (or memory function) by
\begin{eqnarray}
m\frac{\dif C(t)}{\dif t} = - \int_0^t \gamma\left(t-\tau\right)C\left(\tau\right) \dif\tau~,
\label{eq:gle}
\end{eqnarray}
where $m$ is the mass of the molecule, and $C(t)$ is the velocity
autocorrelation (VACF),
\begin{equation}
	C(t) = \left\langle \vec{v}\left(t\right)\cdot 
\vec{v}\left(0\right)\right\rangle/\left\langle v^2\right\rangle~.
\end{equation} 
The friction kernel $\gamma\left(t\right)$ is the autocorrelation
function of the \emph{random} forces on a
molecule.\cite{forster1975hydrodynamic} The standard formality for
extracting $\gamma\left(t\right)$ utilizes Laplace transforms. But
inverting the Laplace transform is non-trivial and we have found the
well-known Stehfest algorithm \cite{Stehfest:1970vj} to be problematic. 
 Berne and Harp \cite{berne1970calculation} developed a
finite-difference-in-time procedure for extracting
$\gamma\left(t\right)$ from Eq.~\eqref{eq:gle}. That procedure is
satisfactory, but sensitive to time resolution in the discrete numerical
$C\left(t\right)$ used as input.  An alternative\cite{you2014role} expresses the Laplace
transform as Fourier integrals, utilizing specifically the transforms
\begin{subequations}
\label{eq:ft}
\begin{align}
\label{eq:ps} \hat{C}'\left(\omega\right) & =  \int_0^\infty C(t) \cos \left(\omega t\right) \dif t~, \\
\hat{C}''\left(\omega\right) & =  \int_0^\infty C(t) \sin \left(\omega t\right) \dif t ~.
\end{align}
\end{subequations}
Then
\begin{eqnarray}
\int_0^\infty \gamma(t) \cos \left(\omega t\right) \dif t = 
\frac{m \hat{C}'\left(\omega\right) }{\hat{C}'\left(\omega\right)^2 + 
\hat{C}''\left(\omega\right)^2}~.
\end{eqnarray}
Taking $\gamma\left(t\right)$ to be even time, the cosine transform is
straightforwardly inverted. $\Omega^2 = \left\langle F^2\right
\rangle/3mk_{\mathrm{B}}T,$ with  $F=\vert \vec{F}\vert$ the force on
the molecule, provides the normalization $\gamma(0) = m\Omega^2.$ A
comparison of these methods are provided in the supplementary information
and Ref.~\citenum{you2014role}.

\section{\label{sec:level2}Results}

\begin{figure*}
\includegraphics[width=7in]{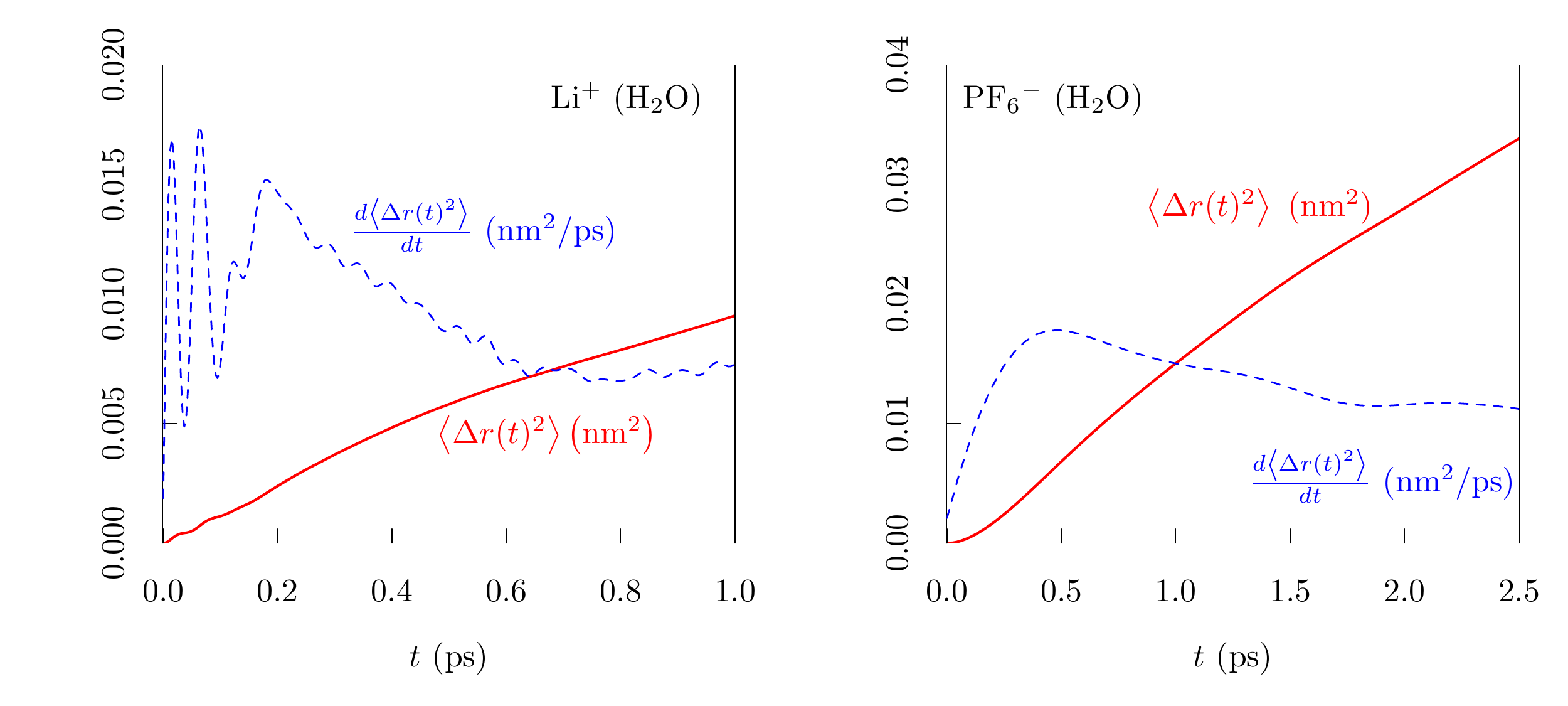}
\caption{Comparison for Li$^+$(aq) (left) and PF$_6{}^-$(aq) (right) of
the mean-squared-displacement (red) and its derivative (blue) with time.
 Oscillatory behavior for Li$^+$ is prominent, not troublesome, and not
evident in the corresponding results for PF$_6{}^-$.}
\label{fig:msd_compare_water} \end{figure*}

\begin{figure} 
\includegraphics[width=3.5in]{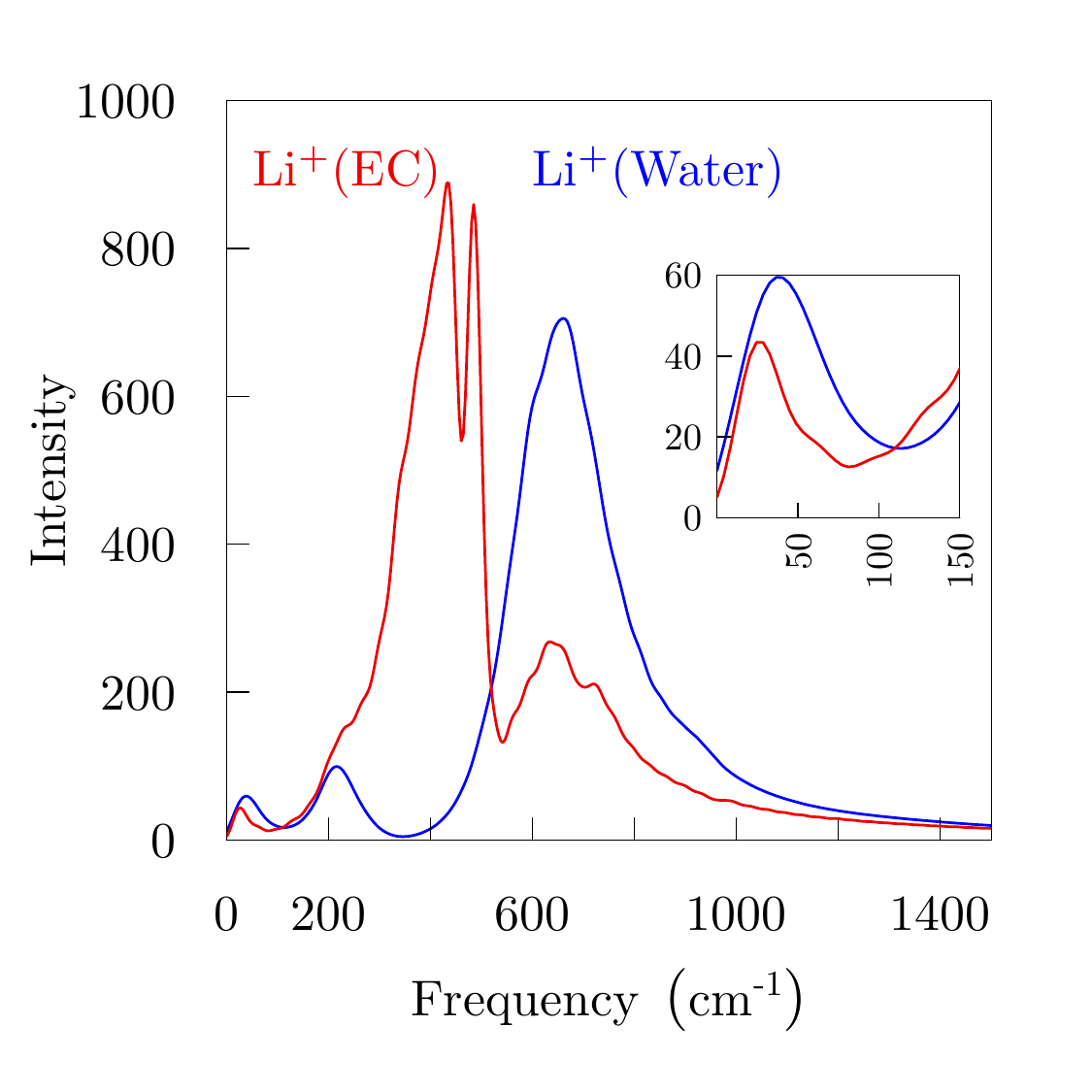} 
\caption{Power spectra (Eq.~\eqref{eq:ps}) of Li$^+$ using the Gromacs
\textit{velacc}\cite{Gromacs_Manual} utility. Red: 1M LiPF$_6$ in EC. To
identify the predominant modes, electronic structure calculations using
Gaussian09\cite{g09} software were performed with the {\tt{b3lyp}}
exchange-correlation density functional and {\tt{6-31+g(d,p)}} basis
set. The frequency mode near 400 cm$^{-1}$ corresponds to motion of a
Li$^+$ ion trapped in a cage formed by its neighbors. The higher
frequency band (near 650 cm$^{-1}$) corresponds to Li$^+$ ion picking up
the scissoring motion of a neighboring carbonate group. Blue: 1M
LiPF$_6$ in Water. Here, the frequency band (near 650 cm$^{-1}$)
corresponds to motion of a Li$^+$ ion trapped in a
cage formed by neighboring water molecules.} \label{fig:spectra} \end{figure}

We discuss quantitative simulation results that lay a basis for
molecule-specific theory of the friction coefficients of ions in
solution. Our initial discussion focuses on dynamics of ions such as
Li$^+$ and PF$_6{}^-$ in water, followed by overall comparisons with
common non-aqueous solvents.
 
\subsection{Oscillatory behavior of Li$^+$ dynamics} 
\begin{figure}
\includegraphics[width=3.5in]{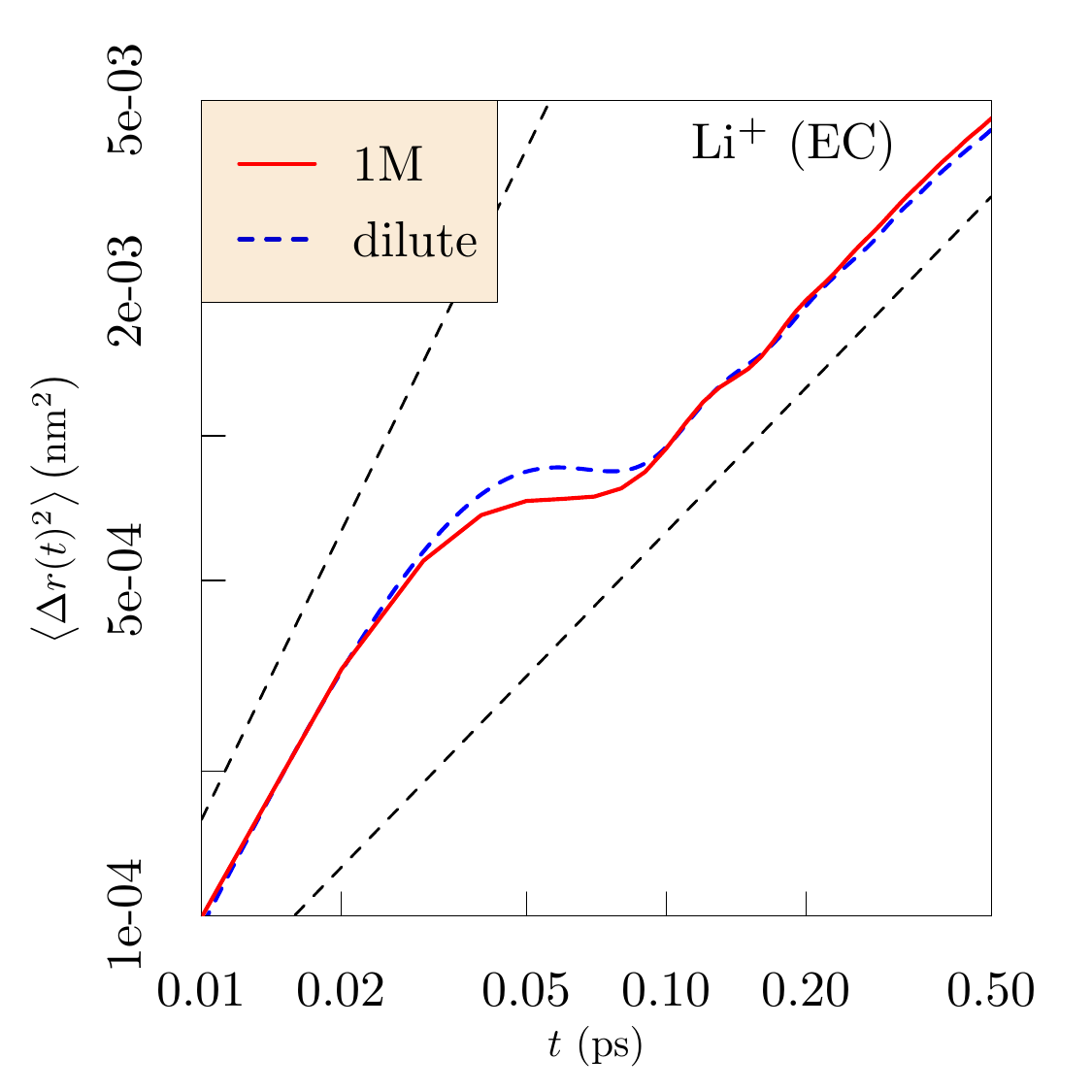}
\caption{The mean-squared-displacement for Li$^+$ ion in EC.  The
black dashed lines indicate slopes of initial ballistic and final diffusive 
behaviors.  The asymptotic slope at long times is not significantly affected 
by concentration and the ion-pairing that is exhibited by this system. }
\label{fig:msd_ionpair} \end{figure}

The Li$^+$ ion has
an unusually small mass, and oscillatory behavior of its dynamics at
short times is prominent compared to PF$_6{}^-$. These differences are
reflected in the mean squared displacement
(FIG.~\ref{fig:msd_compare_water}) of these ions in water. This
short-time behavior has been the particular target of the molecular
time-scale generalized Langevin theory.\cite{adelman1980generalized} The
vibrational power spectrum (FIG.~\ref{fig:spectra}) then provides a more
immediate discrimination of the forces on the ions by the different
solvents. Electronic structure calculations identify the high frequency
vibrations that are related to motion of a Li$^+$ trapped within an
inner solvation shell. In the case of Li$^+$(aq), this frequency occurs
at 650~cm$^{-1}$. Nevertheless, the low frequency ($\omega \approx 0$)
diffusive behavior can be only subtly distinct
for different solution cases (FIG.~\ref{fig:spectra}), including electrolyte
concentration (FIG.~\ref{fig:msd_ionpair}).

\begin{figure*}
\includegraphics[width=7in]{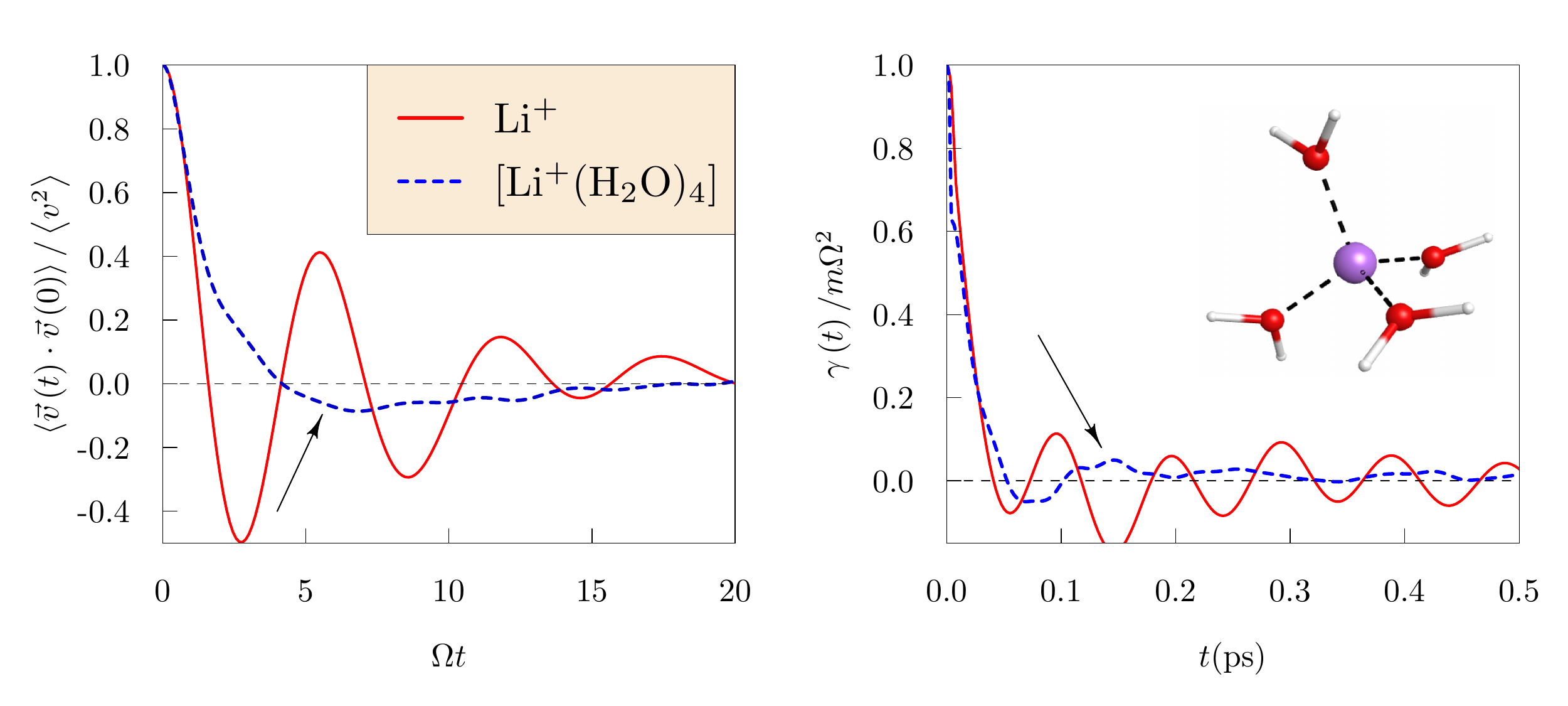}
\caption{Left: Comparison of Li$^+$ in water (red solid line) with
center-of-mass of   
Li$^+\left(\mathrm{H}_2\mathrm{O}\right)_4$ solventberg (blue dashed
line indicated by the arrows). The horizontal axis is made dimensionless with the Einstein
frequency $\Omega \equiv \sqrt{-C^{\prime\prime}\left(t=0\right)}$
evaluated numerically. These values are $\Omega \approx $ 590~cm$^{-1}$
and 116~cm$^{-1}$ for Li$^+$(aq) and the Li$^+$(H$_2$O)$_4$ solventberg,
respectively.  This means that the total temporal extent of the
displayed relaxations are about five times longer for the
Li$^+$(H$_2$O)$_4$ solventberg results than for the Li$^+$(aq). This
time scaling results in matching the initial curvatures of the distinct
functions shown here. The oscillations that are internal to the
solventberg (inset on right) are not reflected in that VACF. The
negative tail relaxation of the solventberg is then qualitatively
similar to that of PF$_6{}^-$ (see Fig.~\ref{fig:PF6_all}). Right:
$\gamma\left(t\right)$ of the solventberg  is also similar to
PF$_6{}^-$(aq) (Fig.~\ref{fig:PF6_all}).} 
\label{fig:solventberg} 
\end{figure*}

\subsection{ Solventberg picture} A common view why the transport
parameters can depend only weakly on the differences in the 
molecular-time-scale dynamics (FIG.~\ref{fig:msd_compare_water})
follows from the appreciation that the exchange
time for inner shell solvent molecules can be long compared to the dynamical
differences. For Li$^+$(aq), that exchange
time is of the order of 30~ps.\cite{friedman1985hydration,Dang:2013fb} Then ion
\emph{plus} inner-shell solvent molecules --- a
\emph{solventberg}\cite{wolynes1978molecular} --- can be viewed as the
transporting species. 

The mean-squared displacement of the ion followed over
times that are long on molecular time-scale 
 but shorter than that exchange time should not differ much from the
mean-squared displacement of the solventberg.  The oscillations internal to
the solventberg, which are reflected in the VACF, are not essential to the
transport. Nevertheless, molecular dynamics simulation permits us to check the
VACF of the center-of-mass of the solventberg. This VACF is free of oscillations
and reveals a negative tail relaxation that is qualitatively similar to
PF$_6{}^-$ (FIG.~\ref{fig:solventberg}). Indeed, previous calculations, treating
both water\cite{Annapureddy:2012dv,Annapureddy:2014isa} and
EC,\cite{Chang:2017fd} fixed a Li$^+$ ion coordinate for calculation of the
\emph{force} autocorrelation. Those prior works indeed also observed this
second, longer time-scale relaxation that the present calculations highlight. 

\subsection{ Overall comparisons}
\begin{figure*}
\includegraphics[width=7in]{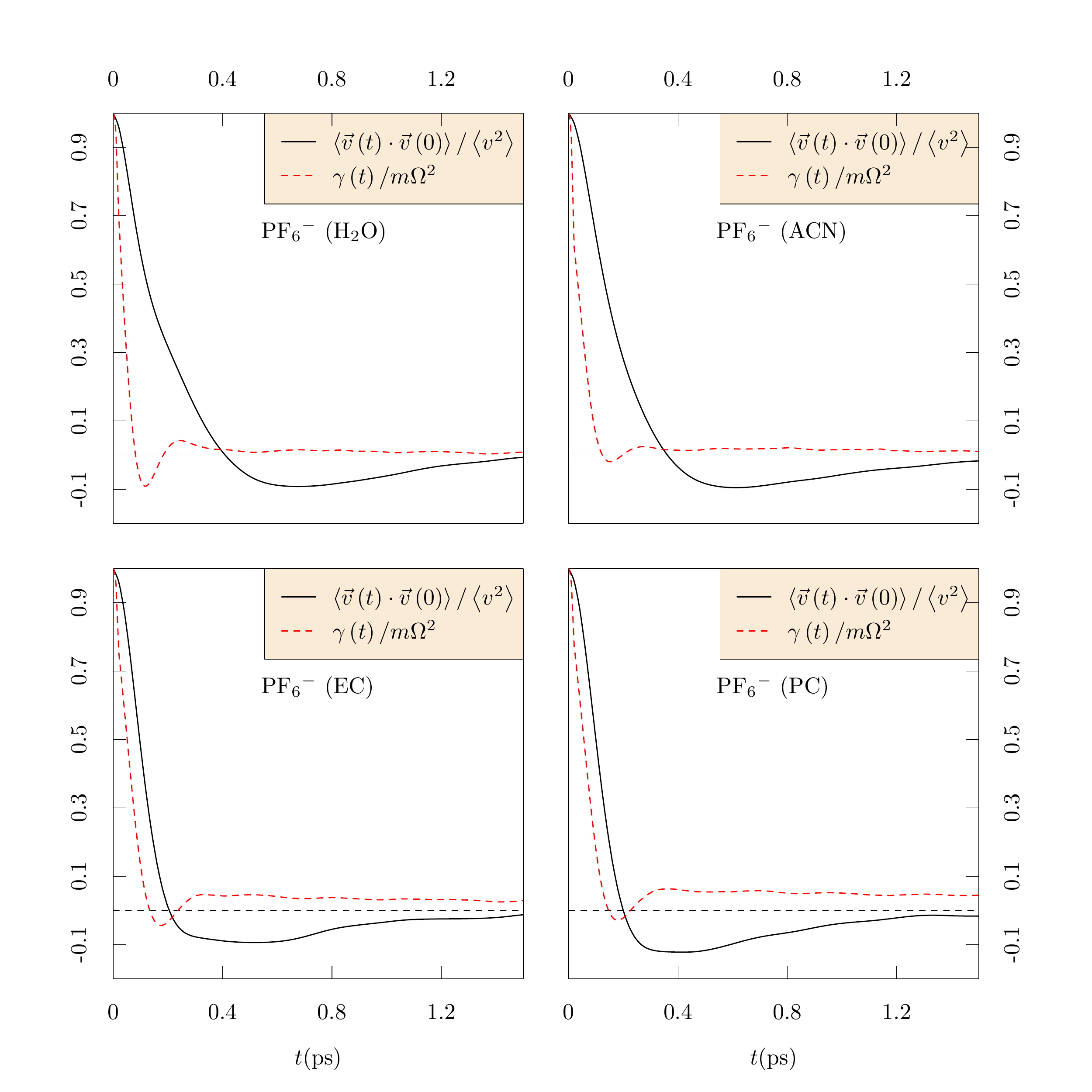}
\caption{The center-of-mass VACF (black solid lines) and the
corresponding friction kernel (red dashed lines), ${\scriptsize \gamma
\left( t \right)}$, for PF$_6{}^-$ from simulations of 1M LiPF$_6$ in
several solvents. While ${\scriptsize \gamma \left( t \right)}$ is
qualitatively similar at short and moderate times, the longer time-scale
relaxation is more prominent for EC and PC compared to water and ACN.
All 1M calculations consisted of 32 Li$^+$ and PF$_6{}^-$ ions together
with the  solvent molecule  numbers that  fix the specified molarity of
the solution.}
\label{fig:PF6_all} \end{figure*}

 The overall comparisons of
these single-ion VACFs and $\gamma\left(t\right)$ for our collection of solvents
(FIG.~\ref{fig:PF6_all}) show these relaxations are similar to each
other for the heavier ion PF$_6{}^-$: a clear second relaxation for $\gamma\left(t\right)$
consistent with 
negative tail relaxations for the VACF. 
This behavior is similar
for other molecular ions considered recently, and including $n$-pentanol as
solvent.\cite{zhu2012pairing}
Numerical VACF results for PF$_6{}^-$(aq) show that the molecular time-scale
relaxation is insensitive to electrolyte concentration and to van~der~Waals
attractive forces (SI).  For Li$^+$, short time-scale oscillatory behavior
masks that longer time-scale relaxation of $\gamma\left(t\right)$, as discussed
above.  Detailed results corresponding to FIG.~\ref{fig:PF6_all} but for a Li$^+$ ion
are provided in the SI.

\section{Conclusions} We extract the VACF and the memory function, $\gamma(t)$,
which characterize the mobility of ions in solution. For the heavier PF$_6{}^-$
ion, velocity relaxations are all similar: negative tail relaxations for the
VACF and a clear second relaxation for $\gamma\left(t\right)$. For the light
Li$^+$ ion, analysis of the solventberg dynamics conform to the standard picture
set by all the PF$_6{}^-$ results. These results lay a quantitative basis for
establishing a molecule-specific theory of the friction coefficients of ions in
solution.

\section{SUPPLEMENTARY MATERIAL} The supplementary material provides a comparison of
methods for extracting the friction kernel, a comparison of Li$^+$
dynamics in different solvents, forcefield parameters for PF$_6{}^-$ and the
effect of removing van der Waals attractions on the dynamics of PF$_6{}^-$(aq).

\section*{Acknowledgement} Sandia National Laboratories is a
multimission laboratory managed and operated by National Technology and
Engineering Solutions of Sandia LLC, a wholly owned subsidiary of
Honeywell International Inc. for the U.S. Department of Energy's
National Nuclear Security Administration under contract DE-NA0003525.
This work is supported by Sandia's LDRD program (MIC and SBR) and by the
National Science Foundation, Grant CHE-1300993. This work was performed,
in part, at the Center for Integrated Nanotechnologies (CINT), an Office
of Science User Facility operated for the U.S. DOE's Office of Science
by Los Alamos National Laboratory (Contract DE-AC52-06NA25296) and SNL.

\clearpage


\begin{thebibliography}{0}%
\makeatletter
\providecommand \@ifxundefined [1]{%
 \@ifx{#1\undefined}
}%
\providecommand \@ifnum [1]{%
 \ifnum #1\expandafter \@firstoftwo
 \else \expandafter \@secondoftwo
 \fi
}%
\providecommand \@ifx [1]{%
 \ifx #1\expandafter \@firstoftwo
 \else \expandafter \@secondoftwo
 \fi
}%
\providecommand \natexlab [1]{#1}%
\providecommand \enquote  [1]{``#1''}%
\providecommand \bibnamefont  [1]{#1}%
\providecommand \bibfnamefont [1]{#1}%
\providecommand \citenamefont [1]{#1}%
\providecommand \href@noop [0]{\@secondoftwo}%
\providecommand \href [0]{\begingroup \@sanitize@url \@href}%
\providecommand \@href[1]{\@@startlink{#1}\@@href}%
\providecommand \@@href[1]{\endgroup#1\@@endlink}%
\providecommand \@sanitize@url [0]{\catcode `\\12\catcode `\$12\catcode
  `\&12\catcode `\#12\catcode `\^12\catcode `\_12\catcode `\%12\relax}%
\providecommand \@@startlink[1]{}%
\providecommand \@@endlink[0]{}%
\providecommand \url  [0]{\begingroup\@sanitize@url \@url }%
\providecommand \@url [1]{\endgroup\@href {#1}{\urlprefix }}%
\providecommand \urlprefix  [0]{URL }%
\providecommand \Eprint [0]{\href }%
\providecommand \doibase [0]{http://dx.doi.org/}%
\providecommand \selectlanguage [0]{\@gobble}%
\providecommand \bibinfo  [0]{\@secondoftwo}%
\providecommand \bibfield  [0]{\@secondoftwo}%
\providecommand \translation [1]{[#1]}%
\providecommand \BibitemOpen [0]{}%
\providecommand \bibitemStop [0]{}%
\providecommand \bibitemNoStop [0]{.\EOS\space}%
\providecommand \EOS [0]{\spacefactor3000\relax}%
\providecommand \BibitemShut  [1]{\csname bibitem#1\endcsname}%
\let\auto@bib@innerbib\@empty
\end{thebibliography}%


\begin{thebibliography}{31}%
\makeatletter
\providecommand \@ifxundefined [1]{%
 \@ifx{#1\undefined}
}%
\providecommand \@ifnum [1]{%
 \ifnum #1\expandafter \@firstoftwo
 \else \expandafter \@secondoftwo
 \fi
}%
\providecommand \@ifx [1]{%
 \ifx #1\expandafter \@firstoftwo
 \else \expandafter \@secondoftwo
 \fi
}%
\providecommand \natexlab [1]{#1}%
\providecommand \enquote  [1]{``#1''}%
\providecommand \bibnamefont  [1]{#1}%
\providecommand \bibfnamefont [1]{#1}%
\providecommand \citenamefont [1]{#1}%
\providecommand \href@noop [0]{\@secondoftwo}%
\providecommand \href [0]{\begingroup \@sanitize@url \@href}%
\providecommand \@href[1]{\@@startlink{#1}\@@href}%
\providecommand \@@href[1]{\endgroup#1\@@endlink}%
\providecommand \@sanitize@url [0]{\catcode `\\12\catcode `\$12\catcode
  `\&12\catcode `\#12\catcode `\^12\catcode `\_12\catcode `\%12\relax}%
\providecommand \@@startlink[1]{}%
\providecommand \@@endlink[0]{}%
\providecommand \url  [0]{\begingroup\@sanitize@url \@url }%
\providecommand \@url [1]{\endgroup\@href {#1}{\urlprefix }}%
\providecommand \urlprefix  [0]{URL }%
\providecommand \Eprint [0]{\href }%
\providecommand \doibase [0]{http://dx.doi.org/}%
\providecommand \selectlanguage [0]{\@gobble}%
\providecommand \bibinfo  [0]{\@secondoftwo}%
\providecommand \bibfield  [0]{\@secondoftwo}%
\providecommand \translation [1]{[#1]}%
\providecommand \BibitemOpen [0]{}%
\providecommand \bibitemStop [0]{}%
\providecommand \bibitemNoStop [0]{.\EOS\space}%
\providecommand \EOS [0]{\spacefactor3000\relax}%
\providecommand \BibitemShut  [1]{\csname bibitem#1\endcsname}%
\let\auto@bib@innerbib\@empty
\bibitem [{\citenamefont {Forster}(1975)}]{forster1975hydrodynamic}%
  \BibitemOpen
  \bibfield  {author} {\bibinfo {author} {\bibfnamefont {D.}~\bibnamefont
  {Forster}},\ }\href@noop {} {\emph {\bibinfo {title} {Hydrodynamic
  fluctuations, broken symmetry, and correlation functions}}},\ \bibinfo
  {series} {Frontiers in Physics}, Vol.~\bibinfo {volume} {47}\ (\bibinfo
  {publisher} {WA Benjamin, Inc.},\ \bibinfo {address} {Reading, Mass.},\
  \bibinfo {year} {1975})\BibitemShut {NoStop}%
\bibitem [{\citenamefont {Zhu}, \citenamefont {Pratt},\ and\ \citenamefont
  {Papadopoulos}(2012)}]{zhu2012pairing}%
  \BibitemOpen
  \bibfield  {author} {\bibinfo {author} {\bibfnamefont {P.}~\bibnamefont
  {Zhu}}, \bibinfo {author} {\bibfnamefont {L.}~\bibnamefont {Pratt}}, \ and\
  \bibinfo {author} {\bibfnamefont {K.}~\bibnamefont {Papadopoulos}},\
  }\href@noop {} {\bibfield  {journal} {\bibinfo  {journal} {J. Chem. Phys.}\
  }\textbf {\bibinfo {volume} {137}},\ \bibinfo {pages} {174501} (\bibinfo
  {year} {2012})}\BibitemShut {NoStop}%
\bibitem [{\citenamefont {Wolynes}(1978)}]{wolynes1978molecular}%
  \BibitemOpen
  \bibfield  {author} {\bibinfo {author} {\bibfnamefont {P.~G.}\ \bibnamefont
  {Wolynes}},\ }\href@noop {} {\bibfield  {journal} {\bibinfo  {journal} {J.
  Chem. Phys.}\ }\textbf {\bibinfo {volume} {68}},\ \bibinfo {pages} {473}
  (\bibinfo {year} {1978})}\BibitemShut {NoStop}%
\bibitem [{\citenamefont {You}, \citenamefont {Pratt},\ and\ \citenamefont
  {Rick}(2014)}]{you2014role}%
  \BibitemOpen
  \bibfield  {author} {\bibinfo {author} {\bibfnamefont {X.}~\bibnamefont
  {You}}, \bibinfo {author} {\bibfnamefont {L.~R.}\ \bibnamefont {Pratt}}, \
  and\ \bibinfo {author} {\bibfnamefont {S.~W.}\ \bibnamefont {Rick}},\
  }\href@noop {} {\bibfield  {journal} {\bibinfo  {journal} {arXiv preprint
  arXiv:1411.1773}\ } (\bibinfo {year} {2014})}\BibitemShut {NoStop}%
\bibitem [{\citenamefont {Zwanzig}\ and\ \citenamefont
  {Bixon}(1970)}]{Zwanzig:1970va}%
  \BibitemOpen
  \bibfield  {author} {\bibinfo {author} {\bibfnamefont {R.}~\bibnamefont
  {Zwanzig}}\ and\ \bibinfo {author} {\bibfnamefont {M.}~\bibnamefont
  {Bixon}},\ }\href@noop {} {\bibfield  {journal} {\bibinfo  {journal} {Phys.
  Rev. A}\ }\textbf {\bibinfo {volume} {2}},\ \bibinfo {pages} {2005} (\bibinfo
  {year} {1970})}\BibitemShut {NoStop}%
\bibitem [{\citenamefont {Metiu}, \citenamefont {Oxtoby},\ and\ \citenamefont
  {Freed}(1977)}]{Metiu:1977wz}%
  \BibitemOpen
  \bibfield  {author} {\bibinfo {author} {\bibfnamefont {H.}~\bibnamefont
  {Metiu}}, \bibinfo {author} {\bibfnamefont {D.~W.}\ \bibnamefont {Oxtoby}}, \
  and\ \bibinfo {author} {\bibfnamefont {K.~F.}\ \bibnamefont {Freed}},\
  }\href@noop {} {\bibfield  {journal} {\bibinfo  {journal} {Phys. Rev. A}\
  }\textbf {\bibinfo {volume} {15}},\ \bibinfo {pages} {361} (\bibinfo {year}
  {1977})}\BibitemShut {NoStop}%
\bibitem [{\citenamefont {Gaskell}\ and\ \citenamefont
  {Miller}(1978)}]{Gaskell:2001cq}%
  \BibitemOpen
  \bibfield  {author} {\bibinfo {author} {\bibfnamefont {T.}~\bibnamefont
  {Gaskell}}\ and\ \bibinfo {author} {\bibfnamefont {S.}~\bibnamefont
  {Miller}},\ }\href@noop {} {\bibfield  {journal} {\bibinfo  {journal} {J.
  Phys. C: Solid State Physics}\ }\textbf {\bibinfo {volume} {11}},\ \bibinfo
  {pages} {3749} (\bibinfo {year} {1978})}\BibitemShut {NoStop}%
\bibitem [{\citenamefont {Balucani}, \citenamefont {Brodholt},\ and\
  \citenamefont {Vallauri}(1999)}]{Balucani:1999ec}%
  \BibitemOpen
  \bibfield  {author} {\bibinfo {author} {\bibfnamefont {U.}~\bibnamefont
  {Balucani}}, \bibinfo {author} {\bibfnamefont {J.~P.}\ \bibnamefont
  {Brodholt}}, \ and\ \bibinfo {author} {\bibfnamefont {R.}~\bibnamefont
  {Vallauri}},\ }\href@noop {} {\bibfield  {journal} {\bibinfo  {journal} {J.
  Phys.: Condensed Matter}\ }\textbf {\bibinfo {volume} {8}},\ \bibinfo {pages}
  {6139} (\bibinfo {year} {1999})}\BibitemShut {NoStop}%
\bibitem [{\citenamefont {Nos{\'e}}(1984)}]{Nose}%
  \BibitemOpen
  \bibfield  {author} {\bibinfo {author} {\bibfnamefont {S.}~\bibnamefont
  {Nos{\'e}}},\ }\href@noop {} {\bibfield  {journal} {\bibinfo  {journal} {Mol.
  Phys.}\ }\textbf {\bibinfo {volume} {52}},\ \bibinfo {pages} {255} (\bibinfo
  {year} {1984})}\BibitemShut {NoStop}%
\bibitem [{\citenamefont {Hoover}(1985)}]{Hoover}%
  \BibitemOpen
  \bibfield  {author} {\bibinfo {author} {\bibfnamefont {W.~G.}\ \bibnamefont
  {Hoover}},\ }\href@noop {} {\bibfield  {journal} {\bibinfo  {journal} {Phys.
  Rev. A}\ }\textbf {\bibinfo {volume} {31}},\ \bibinfo {pages} {1695}
  (\bibinfo {year} {1985})}\BibitemShut {NoStop}%
\bibitem [{\citenamefont {Parrinello}\ and\ \citenamefont
  {Rahman}(1981)}]{barostat}%
  \BibitemOpen
  \bibfield  {author} {\bibinfo {author} {\bibfnamefont {M.}~\bibnamefont
  {Parrinello}}\ and\ \bibinfo {author} {\bibfnamefont {A.}~\bibnamefont
  {Rahman}},\ }\href@noop {} {\bibfield  {journal} {\bibinfo  {journal} {J.
  App. Phys.}\ }\textbf {\bibinfo {volume} {52}},\ \bibinfo {pages} {7182}
  (\bibinfo {year} {1981})}\BibitemShut {NoStop}%
\bibitem [{\citenamefont {Frisch}\ \emph {et~al.}()\citenamefont {Frisch},
  \citenamefont {Trucks}, \citenamefont {Schlegel},\ and\ \citenamefont
  {\emph{et. al.}}}]{g09}%
  \BibitemOpen
  \bibfield  {author} {\bibinfo {author} {\bibfnamefont {M.~J.}\ \bibnamefont
  {Frisch}}, \bibinfo {author} {\bibfnamefont {G.~W.}\ \bibnamefont {Trucks}},
  \bibinfo {author} {\bibfnamefont {H.~B.}\ \bibnamefont {Schlegel}}, \ and\
  \bibinfo {author} {\bibnamefont {\emph{et. al.}}},\ }\href@noop {} {\enquote
  {\bibinfo {title} {Gaussian~09 {R}evision {A}.1},}\ }\bibinfo {note}
  {Gaussian Inc. Wallingford CT 2009}\BibitemShut {NoStop}%
\bibitem [{\citenamefont {Jorgensen}\ and\ \citenamefont
  {Maxwell}(1996)}]{oplsaa}%
  \BibitemOpen
  \bibfield  {author} {\bibinfo {author} {\bibfnamefont {W.~L.}\ \bibnamefont
  {Jorgensen}}\ and\ \bibinfo {author} {\bibfnamefont {D.~S.}\ \bibnamefont
  {Maxwell}},\ }\href@noop {} {\bibfield  {journal} {\bibinfo  {journal} {J.
  Am. Chem. Soc.}\ }\textbf {\bibinfo {volume} {118}},\ \bibinfo {pages}
  {11225} (\bibinfo {year} {1996})}\BibitemShut {NoStop}%
\bibitem [{\citenamefont {Chaudhari}\ \emph {et~al.}(2016)\citenamefont
  {Chaudhari}, \citenamefont {Nair}, \citenamefont {Pratt}, \citenamefont
  {Soto}, \citenamefont {Balbuena},\ and\ \citenamefont
  {Rempe}}]{chaudhari2016scaling}%
  \BibitemOpen
  \bibfield  {author} {\bibinfo {author} {\bibfnamefont {M.~I.}\ \bibnamefont
  {Chaudhari}}, \bibinfo {author} {\bibfnamefont {J.~R.}\ \bibnamefont {Nair}},
  \bibinfo {author} {\bibfnamefont {L.~R.}\ \bibnamefont {Pratt}}, \bibinfo
  {author} {\bibfnamefont {F.~A.}\ \bibnamefont {Soto}}, \bibinfo {author}
  {\bibfnamefont {P.~B.}\ \bibnamefont {Balbuena}}, \ and\ \bibinfo {author}
  {\bibfnamefont {S.~B.}\ \bibnamefont {Rempe}},\ }\href@noop {} {\bibfield
  {journal} {\bibinfo  {journal} {J. Chem. Theory \& Comp.}\ }\textbf {\bibinfo
  {volume} {12}},\ \bibinfo {pages} {5709} (\bibinfo {year}
  {2016})}\BibitemShut {NoStop}%
\bibitem [{\citenamefont {Berendsen}, \citenamefont {Grigera},\ and\
  \citenamefont {Straatsma}(1987)}]{berendsen1987missing}%
  \BibitemOpen
  \bibfield  {author} {\bibinfo {author} {\bibfnamefont {H.}~\bibnamefont
  {Berendsen}}, \bibinfo {author} {\bibfnamefont {J.}~\bibnamefont {Grigera}},
  \ and\ \bibinfo {author} {\bibfnamefont {T.}~\bibnamefont {Straatsma}},\
  }\href@noop {} {\bibfield  {journal} {\bibinfo  {journal} {J. Phys. Chem.}\
  }\textbf {\bibinfo {volume} {91}},\ \bibinfo {pages} {6269} (\bibinfo {year}
  {1987})}\BibitemShut {NoStop}%
\bibitem [{\citenamefont {Soetens}, \citenamefont {Millot},\ and\ \citenamefont
  {Maigret}(1998)}]{soetens1998molecular}%
  \BibitemOpen
  \bibfield  {author} {\bibinfo {author} {\bibfnamefont {J.-C.}\ \bibnamefont
  {Soetens}}, \bibinfo {author} {\bibfnamefont {C.}~\bibnamefont {Millot}}, \
  and\ \bibinfo {author} {\bibfnamefont {B.}~\bibnamefont {Maigret}},\
  }\href@noop {} {\bibfield  {journal} {\bibinfo  {journal} {J. Phys. Chem. A}\
  }\textbf {\bibinfo {volume} {102}},\ \bibinfo {pages} {1055} (\bibinfo {year}
  {1998})}\BibitemShut {NoStop}%
\bibitem [{\citenamefont {Sharma}\ and\ \citenamefont
  {Ghorai}(2016)}]{sharma2016effect}%
  \BibitemOpen
  \bibfield  {author} {\bibinfo {author} {\bibfnamefont {A.}~\bibnamefont
  {Sharma}}\ and\ \bibinfo {author} {\bibfnamefont {P.~K.}\ \bibnamefont
  {Ghorai}},\ }\href@noop {} {\bibfield  {journal} {\bibinfo  {journal} {J.
  Chem. Phys.}\ }\textbf {\bibinfo {volume} {144}},\ \bibinfo {pages} {114505}
  (\bibinfo {year} {2016})}\BibitemShut {NoStop}%
\bibitem [{\citenamefont {Zhu}\ \emph {et~al.}(2011)\citenamefont {Zhu},
  \citenamefont {You}, \citenamefont {Pratt},\ and\ \citenamefont
  {Papadopoulos}}]{zhu2011generalizations}%
  \BibitemOpen
  \bibfield  {author} {\bibinfo {author} {\bibfnamefont {P.}~\bibnamefont
  {Zhu}}, \bibinfo {author} {\bibfnamefont {X.}~\bibnamefont {You}}, \bibinfo
  {author} {\bibfnamefont {L.}~\bibnamefont {Pratt}}, \ and\ \bibinfo {author}
  {\bibfnamefont {K.}~\bibnamefont {Papadopoulos}},\ }\href@noop {} {\bibfield
  {journal} {\bibinfo  {journal} {J. Chem. Phys.}\ }\textbf {\bibinfo {volume}
  {134}},\ \bibinfo {pages} {054502} (\bibinfo {year} {2011})}\BibitemShut
  {NoStop}%
\bibitem [{\citenamefont {Rempe}\ \emph {et~al.}(2000)\citenamefont {Rempe},
  \citenamefont {Pratt}, \citenamefont {Hummer}, \citenamefont {Kress},
  \citenamefont {Martin},\ and\ \citenamefont {Redondo}}]{rempe2000hydration}%
  \BibitemOpen
  \bibfield  {author} {\bibinfo {author} {\bibfnamefont {S.~B.}\ \bibnamefont
  {Rempe}}, \bibinfo {author} {\bibfnamefont {L.~R.}\ \bibnamefont {Pratt}},
  \bibinfo {author} {\bibfnamefont {G.}~\bibnamefont {Hummer}}, \bibinfo
  {author} {\bibfnamefont {J.~D.}\ \bibnamefont {Kress}}, \bibinfo {author}
  {\bibfnamefont {R.~L.}\ \bibnamefont {Martin}}, \ and\ \bibinfo {author}
  {\bibfnamefont {A.}~\bibnamefont {Redondo}},\ }\href@noop {} {\bibfield
  {journal} {\bibinfo  {journal} {J. Am. Chem. Soc.}\ }\textbf {\bibinfo
  {volume} {122}},\ \bibinfo {pages} {966} (\bibinfo {year}
  {2000})}\BibitemShut {NoStop}%
\bibitem [{\citenamefont {Alam}, \citenamefont {Hart},\ and\ \citenamefont
  {Rempe}(2011)}]{alam2011computing}%
  \BibitemOpen
  \bibfield  {author} {\bibinfo {author} {\bibfnamefont {T.~M.}\ \bibnamefont
  {Alam}}, \bibinfo {author} {\bibfnamefont {D.}~\bibnamefont {Hart}}, \ and\
  \bibinfo {author} {\bibfnamefont {S.~L.}\ \bibnamefont {Rempe}},\ }\href@noop
  {} {\bibfield  {journal} {\bibinfo  {journal} {Physical Chemistry Chemical
  Physics}\ }\textbf {\bibinfo {volume} {13}},\ \bibinfo {pages} {13629}
  (\bibinfo {year} {2011})}\BibitemShut {NoStop}%
\bibitem [{\citenamefont {Mason}\ \emph {et~al.}(2015)\citenamefont {Mason},
  \citenamefont {Ansell}, \citenamefont {Neilson},\ and\ \citenamefont
  {Rempe}}]{mason2015neutron}%
  \BibitemOpen
  \bibfield  {author} {\bibinfo {author} {\bibfnamefont {P.}~\bibnamefont
  {Mason}}, \bibinfo {author} {\bibfnamefont {S.}~\bibnamefont {Ansell}},
  \bibinfo {author} {\bibfnamefont {G.}~\bibnamefont {Neilson}}, \ and\
  \bibinfo {author} {\bibfnamefont {S.}~\bibnamefont {Rempe}},\ }\href@noop {}
  {\bibfield  {journal} {\bibinfo  {journal} {J. Phys. Chem. B}\ }\textbf
  {\bibinfo {volume} {119}},\ \bibinfo {pages} {2003} (\bibinfo {year}
  {2015})}\BibitemShut {NoStop}%
\bibitem [{\citenamefont {Stehfest}(1970)}]{Stehfest:1970vj}%
  \BibitemOpen
  \bibfield  {author} {\bibinfo {author} {\bibfnamefont {H.}~\bibnamefont
  {Stehfest}},\ }\href@noop {} {\bibfield  {journal} {\bibinfo  {journal}
  {Comm. ACM}\ }\textbf {\bibinfo {volume} {13}},\ \bibinfo {pages} {47}
  (\bibinfo {year} {1970})}\BibitemShut {NoStop}%
\bibitem [{\citenamefont {Berne}\ and\ \citenamefont
  {Harp}(1970)}]{berne1970calculation}%
  \BibitemOpen
  \bibfield  {author} {\bibinfo {author} {\bibfnamefont {B.~J.}\ \bibnamefont
  {Berne}}\ and\ \bibinfo {author} {\bibfnamefont {G.~D.}\ \bibnamefont
  {Harp}},\ }\href@noop {} {\bibfield  {journal} {\bibinfo  {journal} {Adv.
  Chem. Phys}\ }\textbf {\bibinfo {volume} {17}},\ \bibinfo {pages} {63}
  (\bibinfo {year} {1970})}\BibitemShut {NoStop}%
\bibitem [{\citenamefont {Abraham}\ \emph {et~al.}(2014)\citenamefont
  {Abraham}, \citenamefont {Van Der~Spoel}, \citenamefont {Lindahl},\ and\
  \citenamefont {Hess}}]{Gromacs_Manual}%
  \BibitemOpen
  \bibfield  {author} {\bibinfo {author} {\bibfnamefont {M.}~\bibnamefont
  {Abraham}}, \bibinfo {author} {\bibfnamefont {D.}~\bibnamefont {Van
  Der~Spoel}}, \bibinfo {author} {\bibfnamefont {E.}~\bibnamefont {Lindahl}}, \
  and\ \bibinfo {author} {\bibfnamefont {B.}~\bibnamefont {Hess}},\ }\href@noop
  {} {\enquote {\bibinfo {title} {The {GROMACS} development team {GROMACS} user
  manual version 5.0.4},}\ } (\bibinfo {year} {2014})\BibitemShut {NoStop}%
\bibitem [{\citenamefont {Adelman}(1980)}]{adelman1980generalized}%
  \BibitemOpen
  \bibfield  {author} {\bibinfo {author} {\bibfnamefont {S.}~\bibnamefont
  {Adelman}},\ }\href@noop {} {\bibfield  {journal} {\bibinfo  {journal} {Adv.
  Chem. Phys.}\ }\textbf {\bibinfo {volume} {44}},\ \bibinfo {pages} {143}
  (\bibinfo {year} {1980})}\BibitemShut {NoStop}%
\bibitem [{\citenamefont {Friedman}(1985)}]{friedman1985hydration}%
  \BibitemOpen
  \bibfield  {author} {\bibinfo {author} {\bibfnamefont {H.}~\bibnamefont
  {Friedman}},\ }\href@noop {} {\bibfield  {journal} {\bibinfo  {journal}
  {Chemica Scripta}\ }\textbf {\bibinfo {volume} {25}},\ \bibinfo {pages} {42}
  (\bibinfo {year} {1985})}\BibitemShut {NoStop}%
\bibitem [{\citenamefont {Dang}\ and\ \citenamefont
  {Annapureddy}(2013)}]{Dang:2013fb}%
  \BibitemOpen
  \bibfield  {author} {\bibinfo {author} {\bibfnamefont {L.~X.}\ \bibnamefont
  {Dang}}\ and\ \bibinfo {author} {\bibfnamefont {H.~V.~R.}\ \bibnamefont
  {Annapureddy}},\ }\href@noop {} {\bibfield  {journal} {\bibinfo  {journal}
  {J. Chem. Phys.}\ }\textbf {\bibinfo {volume} {139}},\ \bibinfo {pages}
  {084506} (\bibinfo {year} {2013})}\BibitemShut {NoStop}%
\bibitem [{\citenamefont {Annapureddy}\ and\ \citenamefont
  {Dang}(2012)}]{Annapureddy:2012dv}%
  \BibitemOpen
  \bibfield  {author} {\bibinfo {author} {\bibfnamefont {H.~V.~R.}\
  \bibnamefont {Annapureddy}}\ and\ \bibinfo {author} {\bibfnamefont {L.~X.}\
  \bibnamefont {Dang}},\ }\href@noop {} {\bibfield  {journal} {\bibinfo
  {journal} {J. Phys. Chem. B}\ }\textbf {\bibinfo {volume} {116}},\ \bibinfo
  {pages} {7492} (\bibinfo {year} {2012})}\BibitemShut {NoStop}%
\bibitem [{\citenamefont {Annapureddy}\ and\ \citenamefont
  {Dang}(2014)}]{Annapureddy:2014isa}%
  \BibitemOpen
  \bibfield  {author} {\bibinfo {author} {\bibfnamefont {H.~V.~R.}\
  \bibnamefont {Annapureddy}}\ and\ \bibinfo {author} {\bibfnamefont {L.~X.}\
  \bibnamefont {Dang}},\ }\href@noop {} {\bibfield  {journal} {\bibinfo
  {journal} {J. Phys. Chem. B}\ }\textbf {\bibinfo {volume} {118}},\ \bibinfo
  {pages} {8917} (\bibinfo {year} {2014})}\BibitemShut {NoStop}%
\bibitem [{\citenamefont {Chang}\ and\ \citenamefont
  {Dang}(2017)}]{Chang:2017fd}%
  \BibitemOpen
  \bibfield  {author} {\bibinfo {author} {\bibfnamefont {T.-M.}\ \bibnamefont
  {Chang}}\ and\ \bibinfo {author} {\bibfnamefont {L.~X.}\ \bibnamefont
  {Dang}},\ }\href@noop {} {\bibfield  {journal} {\bibinfo  {journal} {J. Chem.
  Phys.}\ }\textbf {\bibinfo {volume} {147}},\ \bibinfo {pages} {161709}
  (\bibinfo {year} {2017})}\BibitemShut {NoStop}%
\bibitem [{\citenamefont {Weeks}, \citenamefont {Chandler},\ and\ \citenamefont
  {Andersen}(1971)}]{weeks1971role}%
  \BibitemOpen
  \bibfield  {author} {\bibinfo {author} {\bibfnamefont {J.~D.}\ \bibnamefont
  {Weeks}}, \bibinfo {author} {\bibfnamefont {D.}~\bibnamefont {Chandler}}, \
  and\ \bibinfo {author} {\bibfnamefont {H.~C.}\ \bibnamefont {Andersen}},\
  }\href@noop {} {\bibfield  {journal} {\bibinfo  {journal} {J. Chem. Phys.}\
  }\textbf {\bibinfo {volume} {54}},\ \bibinfo {pages} {5237} (\bibinfo {year}
  {1971})}\BibitemShut {NoStop}%
\end{thebibliography}

%

\end{document}